\documentclass[twocolumn,10pt,reprint,secnumarabic,amssymb,superscriptaddress, nobibnotes,aps, prl,nobibnotes,showpacs]{revtex4-1}
\usepackage{amsmath}                
\usepackage{bm}
\renewcommand{\vec}{\bm}
\usepackage{graphicx}

\begin{document}

\title{Magnetizing a complex plasma without a magnetic field}

\author{H. K\"ahlert}
 \affiliation{Heinrich-Heine-Universit\"at D\"usseldorf, Institut f\"ur Theoretische Physik II: Weiche Materie, Universit\"atsstra\ss{}e 1, 40225 D\"usseldorf, Germany}
\author{J. Carstensen}
 \affiliation{Christian-Albrechts-Universit\"at zu Kiel, Institut f\"ur Experimentelle und Angewandte Physik, Leibnizstra\ss{}e 19, 24098 Kiel, Germany}
\author{M. Bonitz}
\email{bonitz@physik.uni-kiel.de}
 \affiliation{Christian-Albrechts-Universit\"at zu Kiel, Institut f\"ur Theoretische Physik und Astrophysik, Leibnizstra\ss{}e 15, 24098 Kiel, Germany}
\author{H. L\"owen}
 \affiliation{Heinrich-Heine-Universit\"at D\"usseldorf, Institut f\"ur Theoretische Physik II: Weiche Materie, Universit\"atsstra\ss{}e 1, 40225 D\"usseldorf, Germany}
\author{F. Greiner}
\author{A. Piel}
 \affiliation{Christian-Albrechts-Universit\"at zu Kiel, Institut f\"ur Experimentelle und Angewandte Physik, Leibnizstra\ss{}e 19, 24098 Kiel, Germany}

\date{\today}

\begin{abstract}
We propose and demonstrate a concept that mimics the magnetization of the heavy dust particles in a complex plasma while leaving the properties of the light species practically unaffected. It makes use of the frictional coupling between a complex plasma and the neutral gas, which allows to transfer angular momentum from a rotating gas column to a well-controlled rotation of the dust cloud. This induces a Coriolis force that acts exactly as the Lorentz force in a magnetic field. Experimental normal mode measurements for a small dust cluster with four particles show excellent agreement with theoretical predictions for a magnetized plasma.
\end{abstract}

\maketitle

Strongly coupled Coulomb systems have very unusual properties including spontaneous spatial ordering and the formation of liquids or even crystals. Coulomb crystals and liquids, originally predicted by Wigner~\cite{wigner_34}, were eventually observed on the surface of helium droplets~\cite{grimes_prl79}, in ion traps \cite{wineland_prl87}, and in complex plasmas~\cite{hayashi_jjapl94, thomas_prl94}. They are also believed to occur in semiconductor quantum dots and quantum wells \cite{reimann_rmp02}, as well as in white dwarf and neutron stars~\cite{fortov2009}. Of particular current interest is their behavior in a magnetic field, where strongly modified oscillation spectra~\cite{uchida_prl04,jiang2007, bonitz_prl10,ott2012} or anomalous diffusion properties~\cite{ott_prl11} have been predicted, and even applications to verify enhanced nuclear reaction rates have been demonstrated~\cite{anderegg2009}. Especially in neutron stars~\cite{potekhin2010}, giant magnetic fields are present that considerably modify the properties of the liquid and crystal states in the outer layers and alter the whole evolution of the star. 

\begin{figure}[t]
\centering
\includegraphics{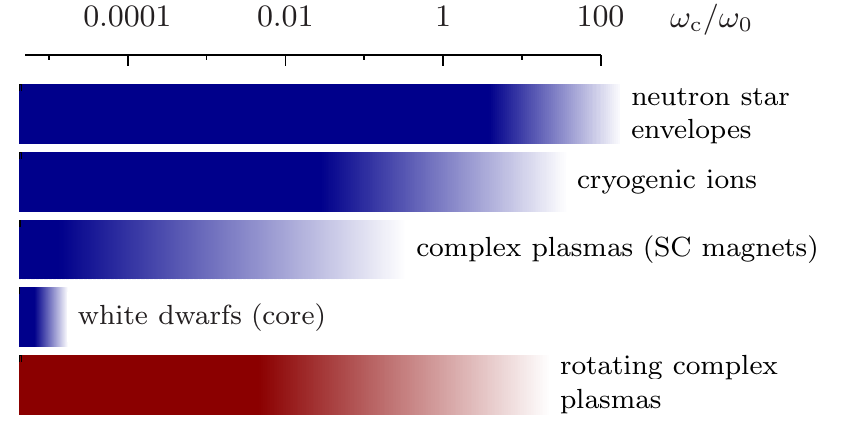}
\caption{(Color online) Estimates of $\omega_\text{c}/\omega_0$ in various strongly coupled plasmas: white dwarf stars~\cite{fortov2009} ($\rho_\text{m}\gtrsim  10^9\,\text{kg m}^{-3}$, $B \lesssim 10^5\,\text{T}$), neutron stars~\cite{potekhin2010} ($\rho_\text{m}\sim 10^7\dots 10^{14}\,\text{kg m}^{-3}$, $B\lesssim 10^{11}\,\text{T}$), ions in Penning traps~\cite{jensen2005,anderegg2009} ($n\sim 10^{7}\dots 10^{8}\,\text{cm}^{-3}$, fields of several Tesla), and complex plasmas~\cite{uchida_prl04,bonitz_prl10,ott2012}. The method presented in this paper (rotating complex plasmas) covers a wide range of magnetizations at room temperature.}\label{fig:overview}
\end{figure}
In a strongly coupled plasma (SCP), the Coulomb interaction energy of two particles, $Q^2/(4\pi\epsilon_0 a)$ [charge $Q$, typical inter-particle distance $a$], is much larger than their thermal energy, $k_BT$. In a {\em magnetized SCP} the particles are, in addition to the electrostatic interactions, subject to the Lorentz force. The {\em degree of magnetization} can be measured by comparing the relevant time scales associated with these two forces~\footnote{Other parameters based on the comparison of length scales (e.g., the cyclotron radius and the Landau length~\cite{anderegg2009}) are also being used.}: While the Coulomb interactions lead to a characteristic vibration frequency~\cite{uchida_prl04} $\omega_0=\sqrt{Q^2/(4\pi \epsilon_0\,m a^3)}$, the cyclotron frequency $\omega_\text{c}=QB/m$ is the relevant parameter for the Lorentz force. For a given magnetic field $B$ and mass density of the plasma $\rho_\text{m}=mn$ (particle mass $m$, number density $n$), the ratio of the two becomes $\omega_\text{c}/\omega_0 = B \sqrt{3\epsilon_0/ \rho_\text{m}}$. Estimates for various strongly coupled astrophysical~\cite{fortov2009, potekhin2010} and laboratory plasmas~\cite{jensen2005,anderegg2009} are presented in Fig.~\ref{fig:overview}.

Complex (dusty) plasmas ~\cite{shukla_rmp2009} have, in recent years, become a prototypical system to study strong correlation effects in {\em unmagnetized} Coulomb systems. They contain highly charged, micrometer-sized particles embedded in a partially ionized electron-ion plasma. Complex plasmas are found in numerous space environments including interstellar clouds, cometary tails, or planetary rings~\cite{goertz1989}. In laboratory experiments, the large particle size and mass make it possible to follow individual particle trajectories with unprecedented spatial and temporal resolution~\cite{morfill2009,rubin2006,bonitz_rop} providing valuable insight into strong coupling phenomena. However, until now, it has not been possible to extend this analysis to magnetized Coulomb systems. The large particle mass (via large $\rho_{\rm m}$)
 limits $\omega_\text{c}/\omega_0$ to values below $0.1\dots 0.5$, even if superconducting magnets and particles with sub-micron diameter are used~\cite{uchida_prl04,bonitz_prl10}, see Fig.~\ref{fig:overview}. Also, inevitably, such fields radically alter the plasma conditions and may lead to filamentation~\cite{schwabe2011}. 

Here, we propose a completely different approach to achieve very strong magnetization effects that overcomes all these limitations. It can be used with standard (micron-sized) particles, does not significantly alter the plasma properties, and does not require expensive magnets at all. The basic idea to effectively ``magnetize'' the complex plasma is to impose the global motion of a rotating neutral gas column on the dust particles, which can be achieved, e.g., by a rotating electrode~\cite{carstensen2009}. In the frame co-rotating with the neutral gas, the Coriolis force takes over the role of the Lorentz force and mimics the effect of a magnetic field, while the centrifugal force renormalizes the confinement. In all other aspects, the equations of motion are equivalent to those in a gas at rest. Related methods, where rotation is induced by different means, have been applied in the context of cold quantum gases~\cite{madison2000,rosenbusch2002,zwierlein2005,fetter2009}.

In the following, we first present the theory underlying our concept and work out the scaling relations for the plasma parameters in a two-dimensional harmonic trap. We then perform a proof-of-principle experiment for the normal modes of a small cluster and find excellent agreement with the theoretical predictions.

In our model, the neutral gas is assumed to rotate as a rigid body with a velocity profile $\vec u(\vec r)= (\Omega\, \hat{\vec e}_z)\times\vec r$. Here, $\hat {\vec e}_z$ denotes the unit vector in the $z$-direction and $\Omega$ the rotation frequency. Particle confinement is provided by a harmonic trapping potential $ V(\rho,z)=\frac{m}{2}\left(\omega_\perp^2 \rho^2 + \omega_z^2 z^2 \right)$, where $\rho=\sqrt{x^2+y^2}$. The equations of motion for the dust particles then have the form ($i=1,\dots, N$)
\begin{align*}
m\ddot{\vec r}_i = &-\nabla_i V(\rho_i,z_i) + \sum_{j\ne i}^N \vec F^\text{int}_{ij} -\nu m\left[ \dot{\vec r}_i - \vec u({\vec r}_i) \right]+ \vec f_i.
\end{align*}
Here, $\vec F^\text{int}_{ij}=-\nabla_i \phi(\rho_{ij},z_{ij})$ denotes the interaction force [$z_{ij}=z_i-z_j$, $\rho_{ij}=\sqrt{x_{ij}^2+y_{ij}^2}$], $\phi(\rho,z)$ the associated potential~\footnote{This form includes ion-wake potentials.}, $\nu$ the dust-neutral friction coefficient, and $\vec f_i(t)$ the stochastic force. The latter has zero mean and the correlation $\langle f_i^\alpha(t) f_j^\beta(t') \rangle = 2m\nu k_\text{B} T_\text{n} \delta_{ij} \delta^{\alpha\beta} \delta(t-t')$, where $i,j\in\{1,\dots,N\}$ and $\alpha,\beta\in\{x,y,z\}$. The neutral gas temperature is denoted by $T_\text{n}$.

In a frame rotating around the $z$-axis at the angular velocity $\Omega$ of the neutral gas, the equations of motion take a form that illustrates the influence of the gas flow.  Denoting the coordinates in the rotating frame with an overbar, the relation between the laboratory frame and the rotating frame reads $\vec r(t)=\vec R(t)\, \bar {\vec r}(t)$, where $\vec R(t)$ is the matrix for rotations around the $z$-axis~(see, e.g., Ref.~\cite{hestenes1999}). Replacing the coordinates leads to
\begin{align*}
 m\ddot{\bar{\vec r}}_i =- \bar\nabla_i \bar V(\bar\rho_i,\bar z_i) + \sum_{j\ne i}^N \bar{\vec F}^\text{int}_{ij} + \bar{\vec F}_\text{Cor}(\dot{\bar{{\vec r}}}_i)-\nu m \dot{\bar{\vec r}}_i + \bar{\vec f}_i,
\end{align*}
where the effective confinement potential is now given by $\bar V(\bar\rho,\bar z)=\frac{m}{2}\left[\left(\omega_\perp^2-\Omega^2\right) \bar\rho^2 + \omega_z^2 \bar z^2 \right]$.

The first observation is that, in the rotating frame, the effective confinement frequency $\bar\omega_\perp=\sqrt{\omega_\perp^2-\Omega^2}$ in the direction perpendicular to the rotation axis is being reduced as a consequence of the centrifugal force. Second, the rotation induces the Coriolis force
\begin{align*}
 \bar{\vec F}_\text{Cor}(\dot{\bar{\vec r}}) =  m \dot{\bar{\vec r}}\times (2\,\Omega\,\hat{\vec e}_z),
\end{align*}
which has the same form as the Lorentz force for a homogeneous magnetic field $\vec B_\text{eff}=(2m\Omega/Q) \hat{\vec e}_z$. The doubled rotation frequency can be identified with the cyclotron frequency, $\omega_\text{c}=2\Omega$. The neutral gas flow velocity $\vec u(\vec r)$ no longer appears explicitly because the coordinate frame rotates at the same angular velocity as the gas. While the form of the inter-particle forces remains unaffected, $\bar{\vec F}^\text{int}_{ij}=-  \bar\nabla_i\phi(\bar\rho_{ij},\bar z_{ij})$, the components of the random force occur in a mixed form, $\bar{\vec f}_i(t)=\vec R^T(t){\vec f}_i(t)$. However, its statistical properties are the same as in the laboratory frame, i.e., a Gaussian white noise is retained, $\langle \bar{\vec f}_i(t) \rangle=0$, and $\langle \bar{f}_i^\alpha(t) \bar{f}_j^\beta(t') \rangle = 2m\nu k_\text{B} T_\text{n} \delta_{ij} \delta^{\alpha\beta} \delta(t-t')$.

Let us now estimate the effective ``magnetic fields'' $|\vec{B}_\text{eff}|=2 m \Omega/Q$ that can be reached. With rotation frequencies of $\Omega\approx10\,\text{Hz}$, a charge of $|Q|\approx 10^{4}e$, and a mass of $m\approx 10^{-12}\,\text{kg}$, one easily generates effective magnetic fields exceeding $10^4\,\text{T}$, which is far beyond the capabilities of superconducting magnets. At very high rotation speeds, the centrifugal force weakens the horizontal confinement~\cite{carstensen2010}, which may give rise to a deformed plasma shape and a lower dust density. However, especially in two-dimensional systems~\cite{feng2008}, the shape of the dust monolayer will remain unaffected, and a lower density could even be beneficial as it decreases $\omega_0$, thus allowing to reach even larger magnetizations $2\Omega/\omega_0$. 

To illustrate these effects in more detail, we consider the limit $\omega_z\gg \omega_\perp$. This situation is typically encountered in dusty plasma experiments in radio-frequency discharges, where the particles are located in the sheath region above the lower electrode. Their interaction is well described by the Yukawa potential $\phi(r)=(Q^2/4\pi\epsilon_0\,r)\exp(-r/\lambda)$ with the Debye screening length $\lambda$. As a typical inter-particle distance we choose $a(\Omega)=[Q^2/(4\pi\epsilon_0\,m\bar\omega_\perp^2)]^{1/3}$, which is equivalent to (up to a numerical factor) the ground state distance of two trapped particles with Coulomb interaction. The effective trap frequency $\bar{\omega}_\perp$ then coincides with the previous definition of $\omega_0=\sqrt{Q^2/(4\pi \epsilon_0\,m a^3)}$ and provides a natural frequency unit for particle vibrations. Both the distance and time scales,  $a(\Omega)$ and $\bar\omega_\perp^{-1}(\Omega)$, increase upon faster rotation. Since the electron and ion time scales are orders of magnitude larger than the rotation frequencies, the charge $Q$ and the Debye length $\lambda$ (i.e., the plasma properties) remain constant to high accuracy.

The problem is then governed by the following dimensionless parameters: the coupling parameter $\Gamma(\Omega)=Q^2/(4\pi\epsilon_0 a(\Omega)\,k_\text{B} T_\text{n})$, the screening parameter $\kappa(\Omega)=a(\Omega)/\lambda$, the damping rate $\gamma(\Omega)=\nu/\bar\omega_\perp(\Omega)$, and the effective magnetization
\begin{equation}\label{eqn:magnetization}
\frac{2 \Omega}{\bar\omega_\perp(\Omega)}=\frac{2(\Omega/\omega_\perp)}{(1-\Omega^2/\omega_\perp^2)^{1/2}}.
\end{equation}
\begin{figure}
\centering
\includegraphics{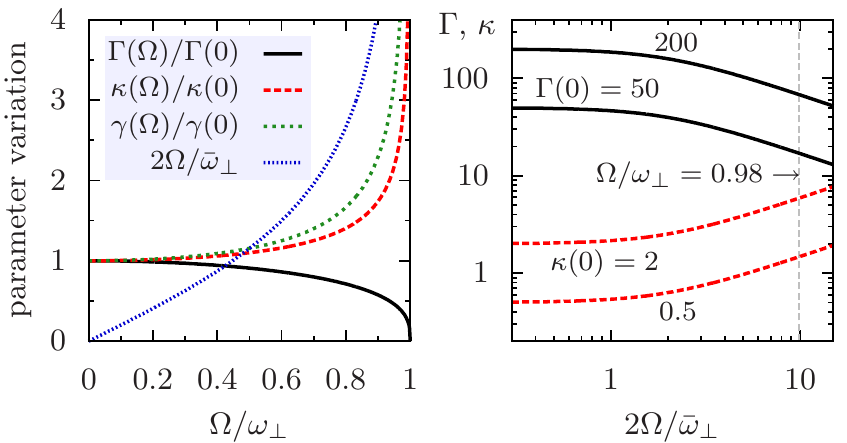}
\caption{(Color online) Left: Variation of the effective magnetization, the coupling parameter, screening parameter and damping constant with the rotation frequency in a finite two-dimensional system. Right: Scaling of $\Gamma$ and $\kappa$ as a function of the effective magnetization $2\Omega/\bar{\omega}_\perp$ for typical complex plasma conditions.}\label{fig:scaling}
\end{figure}

It is obvious that, already for a relatively slow rotation with $\Omega=\omega_\perp/2$, the plasma is strongly ``magnetized'' ($2\Omega/\bar{\omega}_\perp=1.15$), see the left panel of Fig.~\ref{fig:scaling}. The associated changes (compared to the non-rotating system) of the coupling and screening parameter are small. These parameters scale as $\Gamma(\Omega)/\Gamma(0)=(1-\Omega^2/\omega_\perp^2)^{1/3}$ and $\kappa(\Omega)/\kappa(0)=(1-\Omega^2/\omega_\perp^2)^{-1/3}$, respectively, and start to change substantially when the centrifugal force noticeably increases the inter-particle distance ($\Omega/\omega_\perp \gtrsim 0.7$), see the left panel of Fig.~\ref{fig:scaling}. In this regime, the decrease of $\bar\omega_\perp$ is largely responsible for the dramatic increase of the magnetization parameter. We also show the dimensionless damping rate, which scales as $\gamma(\Omega)/\gamma(0)=(1-\Omega^2/\omega_\perp^2)^{-1/2}$. This means that, in experiments, the gas pressure should be sufficiently low allowing for a small neutral gas friction coefficient $\nu$ before start of the rotation. The right panel of Fig.~\ref{fig:scaling} shows $\Gamma$ and $\kappa$ as a function of the effective magnetization in a parameter regime that should be easily accessible in dusty plasma experiments, see below~\footnote{Evidently, the analogy between the rotation and the cyclotron frequency reaches certain limits when plasma confinement is considered. While an increase of $\Omega$ decreases the effective confinement of the dust cloud, a stronger magnetic field gives rise to an improved plasma confinement.}.

\begin{figure}
\centering\includegraphics[width=0.4\textwidth]{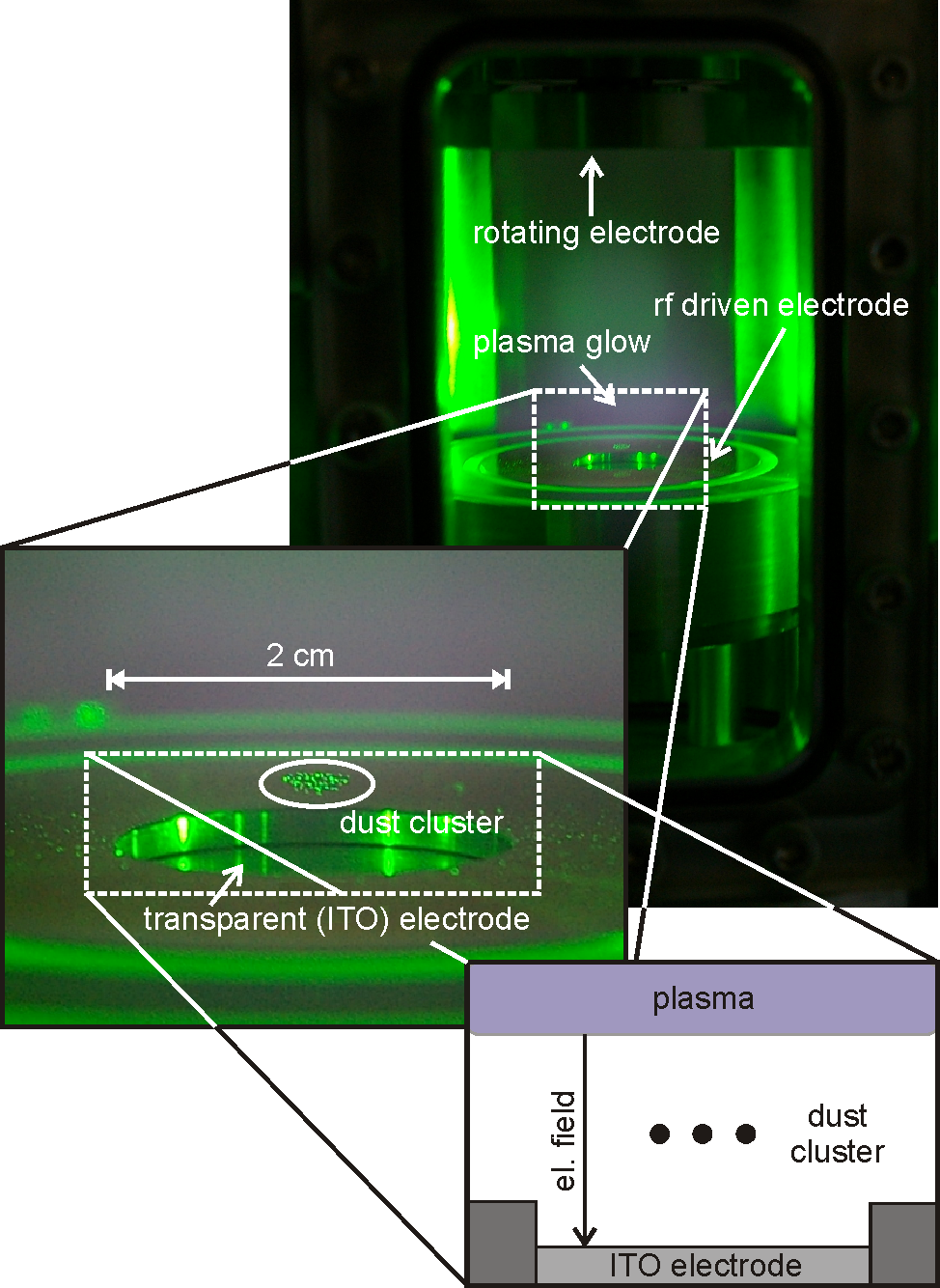}
\caption{(Color online) Experimental setup: Dust particles are confined in a layer approximately 1 cm above the driven electrode. The electric field within the plasma sheath compensates gravity and provides vertical confinement, while horizontal confinement is realized by a cylindrical cavity of 20 mm in diameter and 2 mm in depth. The particles are illuminated from the side by a laser fan and can be observed from the bottom through a transparent electrode coated with indium tin oxide (ITO). The top electrode is rotatable.}\label{fig:setup}
\end{figure}
In the following, we present a proof-of-principle experiment to verify the efficiency of the proposed concept. A sketch of the experimental setup is shown in Fig.~\ref{fig:setup}. The experiments were performed in a 13.56 MHz capacitively coupled radio-frequency discharge at a gas pressure of $p=0.4$ Pa (Argon). Spherical particles with a diameter of $d=21.8$ $\mu$m and a mass of $m=6.46\cdot 10^{-12}\,\text{kg}$ are injected into the plasma, where they form two-dimensional clusters. The upper electrode can be set into rotation with frequencies up to $30\,\text{Hz}$, which causes a vertically sheared rotational motion of the neutral gas column~\cite{carstensen2009}.

\begin{figure}
\centering
\includegraphics[width=0.4\textwidth]{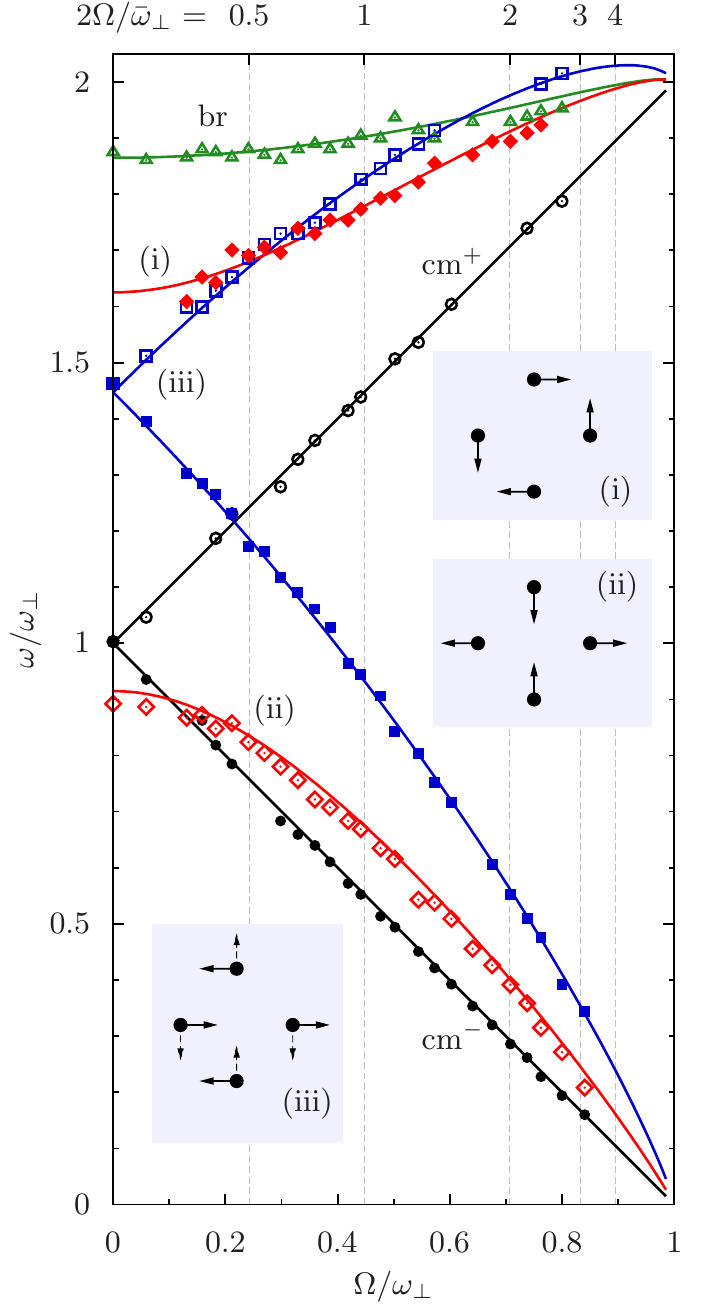}
\caption{(Color online) Comparison of experiment and theory: Frequencies of the seven non-trivial normal modes of $N=4$ dust particles as a function of the scaled rotation frequency. Symbols denote experimental results ($\omega_\perp/2\pi=2.52\,\text{s}^{-1}$), lines are the theoretical eigenfrequencies with $\kappa(0)=0.7$. The eigenvectors in the unmagnetized limit~\cite{amiranashvili2001} ($\Omega=0$) are sketched for four particular modes. Note that the two modes (iii) are degenerate in this case. The breathing mode and the two center-of-mass modes are indicated by ``br'' and ``$\text{cm}^\pm$'', respectively. The effective magnetization $2\Omega/\bar\omega_\perp$ is shown on the upper axis [Eq.~\eqref{eqn:magnetization}].}\label{fig:n=4}
\end{figure}

We concentrate on the dynamics of a small ensemble of $N=4$ particles and analyze their normal modes. The normal modes of small 2D clusters have already been studied experimentally~\cite{melzer2001}, but for magnetized dusty plasmas only theoretical predictions exist~\cite{kong2004}. We first consider the center-of-mass (sloshing) mode. Since the effective confinement in the rotating frame is harmonic, the center-of-mass coordinate $\bar{\vec r}_\text{cm}(t)$ is independent of the interaction (Kohn theorem) and obeys the same equation of motion as a single particle~\cite{jimenez2008}. In the absence of rotation, the two center-of-mass modes are degenerate with $\omega_{\text{cm}}=\omega_\perp$. For $\Omega>0$, however, this degeneracy is lifted, and the frequencies read $\omega_{\text{cm}}^{\pm}=\sqrt{\left(\omega_\text{c}/2\right)^2 + \bar\omega_\perp^2} \pm \omega_\text{c}/2 = \omega_\perp\pm \Omega$. Here, $\bar\omega_\perp=\sqrt{\omega_\perp^2-\Omega^2}$ is the effective trap frequency in the rotating frame. The experimental results obtained from the spectrum of $\bar{\vec r}_\text{cm}(t)$ are depicted in Fig.~\ref{fig:n=4} and show remarkable agreement with the theoretical prediction.

It is crucial for our scheme to further verify the accuracy of the remaining modes, which are sensitive to both the magnetic field \textit{and} inter-particle correlations. By linearizing the equations of motion in the rotating frame, we determine the eigenfrequencies $\omega$ and eigenvectors $\vec v_i$ from~\cite{kong2004} $\left( \omega^2 \delta_{ij}\delta^{\alpha\beta} - \bar H_{ij}^{\alpha\beta}/m -2i\,\omega \Omega\, \delta_{ij}\epsilon^{\alpha \beta z} \right) v_{j}^\beta=0$, where $\alpha,\beta\in\{x,y\}$ and $i,j\in\{1,\dots,N\}$. Further, $\bar H_{ij}^{\alpha\beta}$ denotes the Hessian of the total potential energy in the rotating frame, $\delta_{ij}$ $(\delta^{\alpha\beta})$ the Kronecker delta, and $\epsilon^{\alpha\beta z}$ the Levi-Civita symbol. The cluster configuration for $N=4$ is a square with particles located a distance $R$ from the trap center, which is calculated from Eq.~(6) in Ref.~\cite{amiranashvili2001}. Even though a finite dust-neutral friction parameter is essential to put the plasma into rotation, it is sufficiently low to be negligible for the calculation of the eigenfrequencies ($\nu/\omega_\perp \approx 1/50$). To determine the mode frequencies experimentally, we calculated the projection $P(t)=\sum_{i=1}^N \bar {\vec r}_i(t)\cdot \vec v_i$ of the particle trajectories $\bar {\vec r}_i(t)$ on the eigenvectors $\vec v_i$ of the unmagnetized system, see Fig.~\ref{fig:n=4}. The breathing mode corresponds to a radial expansion and contraction of the cluster. The spectrum of $P(t)$ shows a peak at the associated mode frequency, which can be tracked as the rotation frequency is varied.

The measurements are compared with the theoretical results in Fig.~\ref{fig:n=4}. As for the case of the sloshing modes, we observe excellent agreement with the normal modes of a magnetized plasma. Effective magnetizations $2\Omega/\bar\omega_\perp \gtrsim 3$ allow us to clearly verify the predicted splitting of the normal modes into the upper and lower branch~\cite{kong2004}.

To summarize, we have presented a simple approach to ``magnetize'' a complex plasma. The idea is based on the correspondence between charged particles in a magnetic field and particles in a rotating gas column. The possibility to put the plasma into rotation takes advantage of the dissipative nature of complex plasmas in which the neutral gas acts as a highly effective transmission agent of angular momentum. We demonstrated that, with very modest rotation frequencies applied to only one electrode, strongly correlated particles in a rotating dusty plasma reproduce, to high accuracy, the normal modes of a magnetized system.
Evidently, our concept can also be realized by other means and opens new unique possibilities for highly accurate studies of {\em strongly correlated and strongly magnetized plasmas}. The advantage compared to cryogenic ions is the broad range of accessible plasma parameters that can be varied independently (coupling strength, screening, dissipation) and the availability of single-particle resolution at room temperature conditions. It should be possible to create SCP states with extreme magnetizations potentially even comparable to those exotic ones in the outer layers of neutron stars. While our technique does not require any (superconducting) magnet at all, use of the latter in combination with plasma rotation allows to create novel types of plasmas. In such plasmas, there would be effectively two ``magnetic fields'' that can be controlled independently---one of which (the rotation) affects only the heavy particles whereas the second (``real'') field influences the electrons and ions, allowing for an effective control of the interaction between the dust particles. The combination of a rotating flow with a magnetic field also provides a promising avenue to access magnetorotational instabilities on the particle scale, which are relevant for the understanding of accretion disks~\cite{armitage2011}.

This work was supported by the DFG via SFB~TR6 and SFB~TR24.

%

\end{document}